\documentclass{ws-procs9x6}
\usepackage{graphicx}                   
\input{epsf.sty}                        
\input{psfig.sty}                       

\def\gtrsim{\mathrel{\hbox{\rlap{\hbox{\lower4pt\hbox{$\sim$}}}\hbox{$>$}}}}
\def\lesssim{\mathrel{\hbox{\rlap{\hbox{\lower4pt\hbox{$\sim$}}}\hbox{$<$}}}}

\begin{document}

\title{Solving GRBs and SGRs puzzles by precessing Jets}

\author{D.Fargion$^*$, O.Lanciano, P.Oliva}

\address{Physics Department, University of Rome, "La Sapienza",\\
and INFN, Ple.A.Moro 5, 00185,Rome, Italy\\
$^*$E-mail: ab\_daniele.fargion@roma1.infn.it\\}

\begin{abstract}
A persistent, thin micro-nano sr beamed gamma jet may be ejected
from BH and Pulsars, powered by ultra-relativistic electron pairs.
These precessing and spinning $\gamma$ jet are originated by Inverse
Compton and-or Synchrotron Radiation at pulsars or micro-quasars
sources. They are most powerful at Supernova birth blazing, once on
axis, to us and flashing GRB detector. The trembling of the thin jet
explains naturally the observed erratic multi-explosive structure of
different GRBs. The jets are precessing (by binary companion or
inner asymmetry) and decaying on time scales of a few hours but they
usually keep staying inside the observer cone view only few seconds
duration times (GRB); the jets whole lifetime, while decaying in
output, could survive as long as thousands of years, linking huge
GRB-SN jet apparent Luminosity to more modest SGR relic Jets (at
corresponding X-Ray pulsar output). Therefore long-life SGR may be
repeating and if they are around our galaxy they might be observed
again as the few known ones and the few rarer extragalactic XRFs.
The orientation of the beam respect to the line of sight plays a key
role in differentiating the wide GRB morphology. The relativistic
cone is as small as the inverse of the electron progenitor Lorentz
factor. The hardest and brightest gamma spectra are hidden inside
the inner gamma jet axis. To observe the inner beamed GRB events,
one needs the widest SN sample and the largest cosmic volumes. The
most beamed the hardest. On the contrary, the nearer ones, within
tens Mpc distances, are mostly observable on the cone jet
\textit{periphery}. Their consequent large impact crossing angle
leads to longest \textit{anomalous} SN-GRB duration, with lowest
fluency and the softest spectra, as in earliest GRB98425 and recent
GRB060218 signature. A majority of GRB jet blazing much later
(weeks, months after) may hide their progenitor explosive SN
after-glow and therefore they are called \textit{orphan} GRB. The
late law power GRBs are observable as local SGRs (or in outer near
extragalactic, XRF) and they are linked to anomalous X-ray AXPs.
Conical shape of few nebulae and the precessing jet of few known
micro-quasar, describe in space the model signature. Recent
outstanding episode of X-ray precursor, ten minutes before the main
GRB event, cannot be understood otherwise.
\end{abstract}

\bodymatter

\section{Introduction:  GRB-SGR open questions}
Why GRBs are so spread in their total  energy, (above 6 orders of
magnitude) and in their peak energy  following the so-called Amati
correlation\cite{Amati}? Does the Amati law imply more and more new
GRB families? Why, as shown below the GRB energy is not a constant
but a growing function (almost quadratic) of the red-shift?
\begin{figure}[t]
\begin{center}
\includegraphics[width=3.2in]{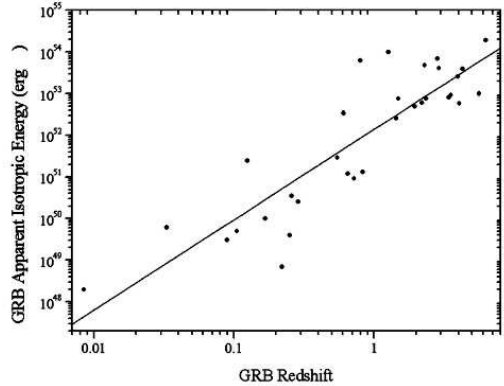}
\caption{Most Recent Swift GRB isotropic apparent energy\cite{Amati}
versus their observed red-shift in Log-Log plot: earlier GRB020903,
GRB030329, GRB031203 and  the nearest GRB980425, at redshift
$z=0.008$, and most distant GRB050904 at $z\simeq6.29$ have been
added. The consequent most correlated law for the apparent energy
growth with distances is nearly quadratic in their red-shift:
$E_{iso}=10^{52.13}\cdot z^{2.147}$ erg. The probability of such a
correlation by chance is below $10^{-4}$. Therefore GRB are
apparently \textit{not standard candle}. This growing energy puzzle
may be solved assuming both threshold cut off for far off-axis GRB
and extreme beaming selectivity at cosmic edges for most energetic
ones.} \label{fig1}
\end{center}
\end{figure}
Why are the harder and more variable GRBs (\cite{Lazzati, Fa99})
found at higher redshifts contrary to expected Hubble law? Why
does the output power of GRB vary in a range (\cite{Fa99}) of 8-9
orders of magnitudes with the most powerful events residing at the
cosmic edges (\cite{Yo2004})? Why has it been possible to find in
the local universe (at distances 40-150 Mpc just a part over a
million of cosmic space) at least two nearby events (GRB980425 at
$z=0.008$ and recent GRB060218 at $z=0.03$)? Most GRBs should be
located at largest volumes, at $z\geq1$ (\cite{Fa99}). Why are
these two nearby GRBs so much under-luminous (\cite{Fa99})? Why
are their evolution times so slow and smooth? Why do their
afterglows show so many bumps and re-brightening as the well-known
third nearest event, GRB030329? Why do not many GRB curves show
monotonic decay (an obvious consequence of a one-shot explosive
event), rather they often show sudden re-brightening or bumpy
afterglows at different time scales and wavelengths (\cite{Stanek,
DaF03}) - see e.g. GRB050502B\cite{Falcone}? Why have there been a
few GRBs and SGRs whose spectra and time structure are almost
identical if their origin is so different (beamed explosion for
GRB versus isotropic magnetar)\cite{Fa99, Woo99}? How can a jetted
fireball (with an opening angle of $5^o$-$10^o$) release a power
nearly 6 orders of magnitude more energetic than the corresponding
isotropic SN?
\begin{figure}[t]
\includegraphics[width=1.3in]{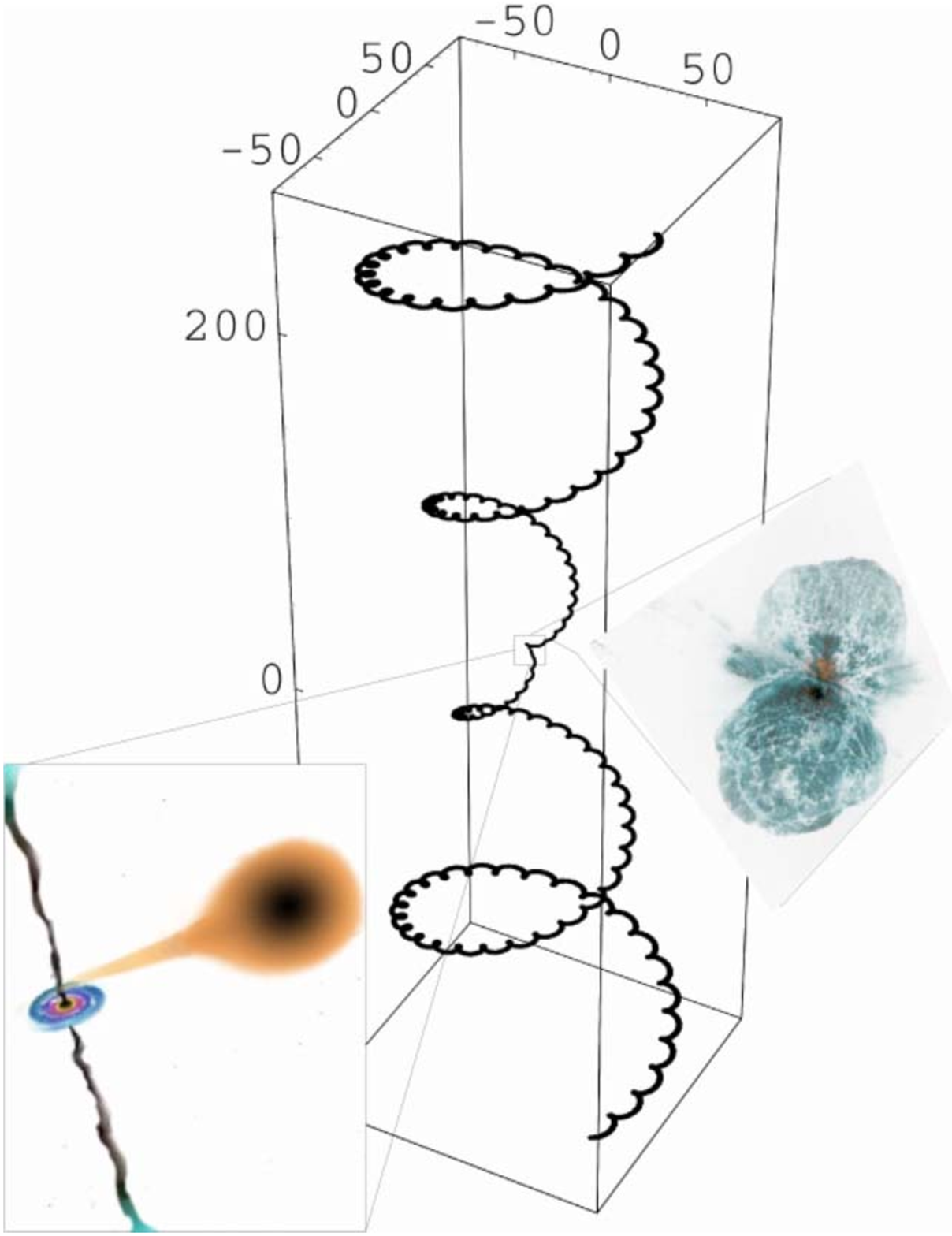}
\includegraphics[width=1.1in]{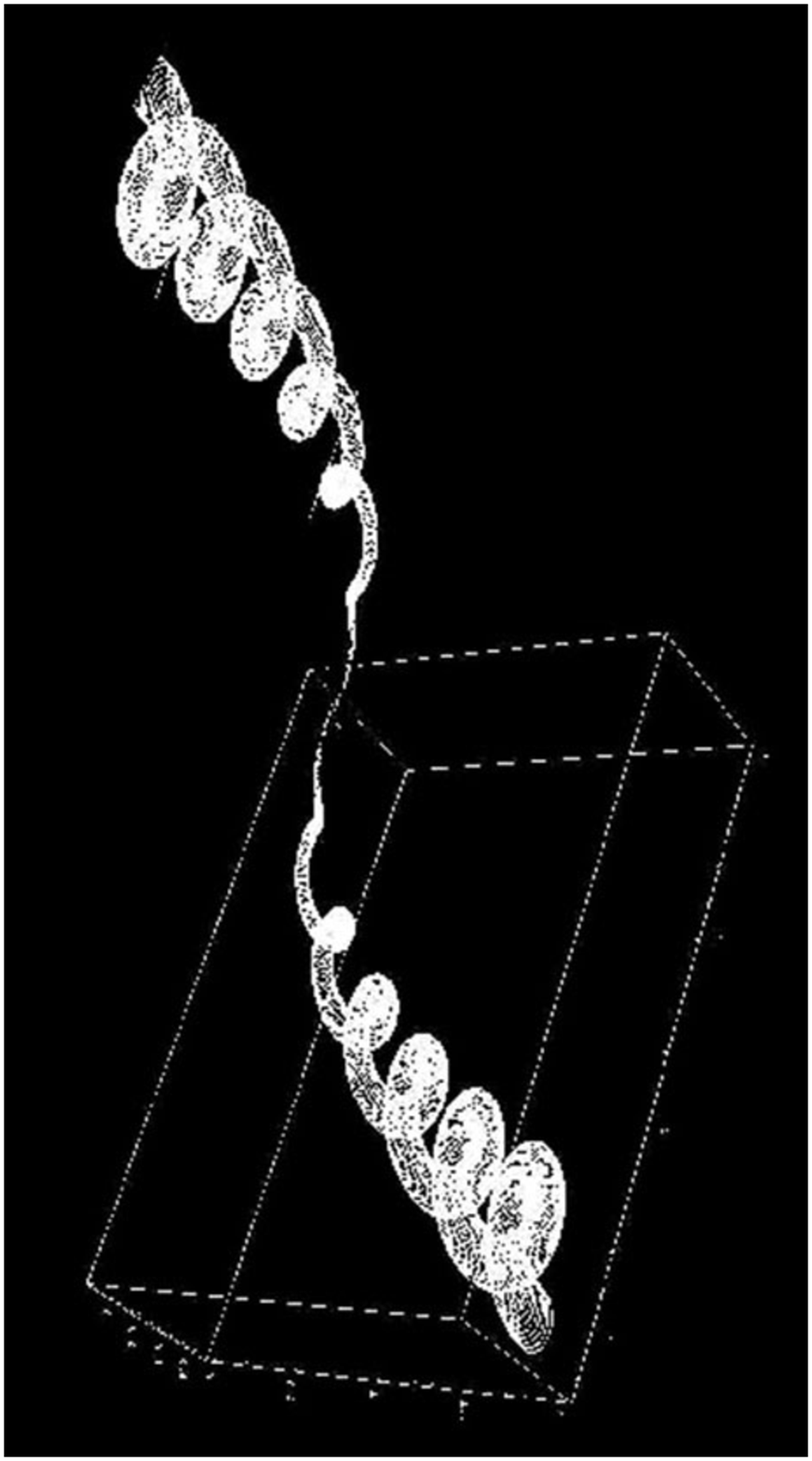}
\includegraphics[width=0.5in]{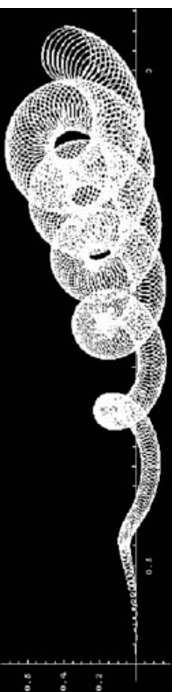}
\includegraphics[width=1.4in]{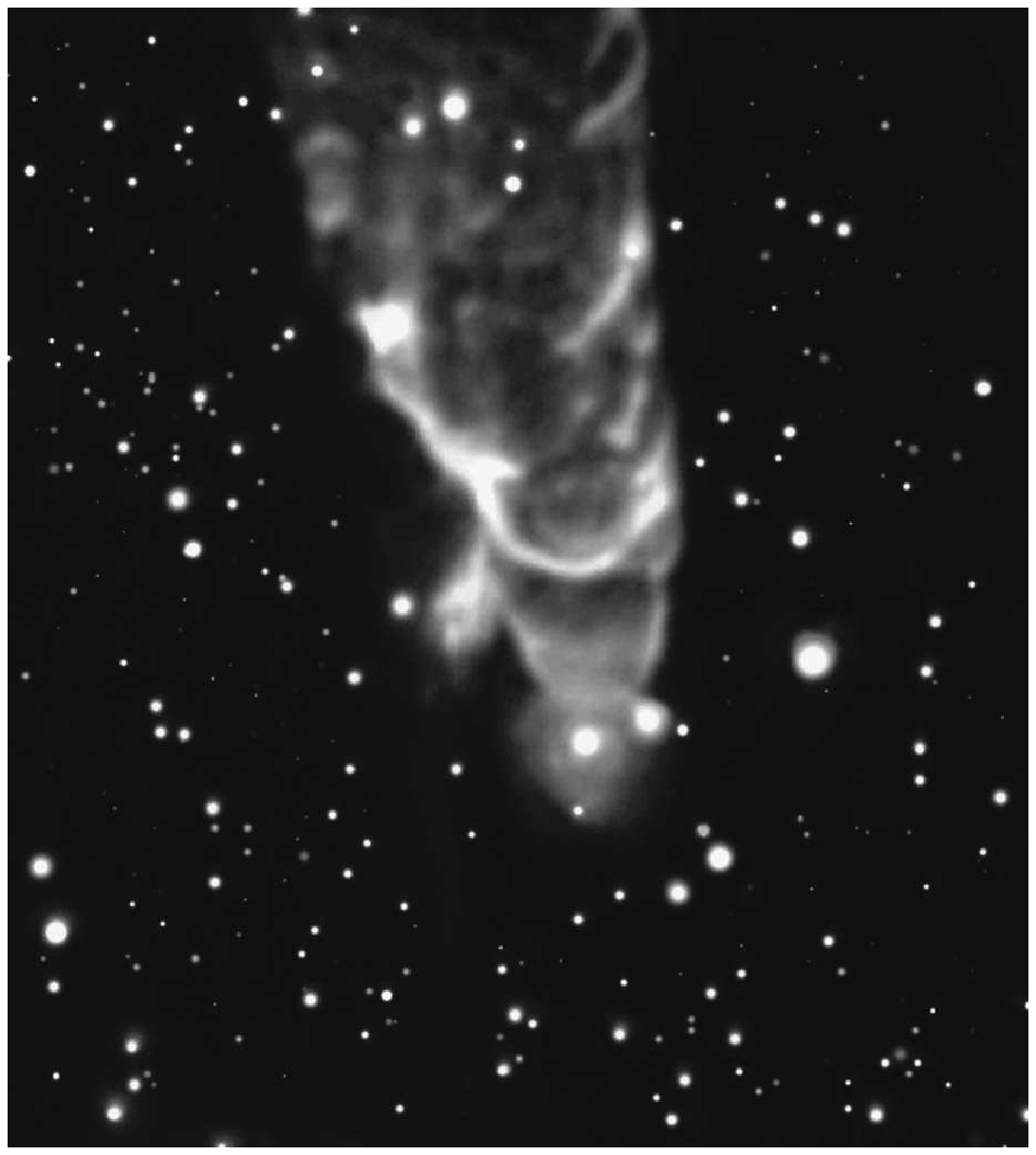}
\caption{From the left to the right: A possible \emph{3D} structure
view of the precessing jet obtained with, for instance, a non linear
precessing, while spinning, gamma jet; at its center the "explosive"
SN-like event for a GRB  or a steady binary system for a SGRs where
an accretion disc around a compact object, powers a collimated
precessing jet. In the two center figures, the \emph{3D} and the
projected \emph{2D} of such  similar precessing Jet. In the right
last panel we show an Herbig Haro-like object HH49, whose spiral
jets are describing, at a lower energy scale, the ones in
micro-quasars such as well known SS-433.} \label{fig2}
\end{figure}
How can re-brightening take place in the  X-ray  and optical
afterglows (\cite{DaF03})? How can some ($\sim6\%$) of the  GRBs
(or a few SGRs) survive the "tiny" (but still extremely powerful)
explosion of its \textit{precursor} without any consequences, and
then explode catastrophically few minutes later? In such a
scenario, how could the very recent GRB060124 (at redshift
$z=2.3$) be preceded by a 10 minutes precursor, and then being
able to produce multiple bursts hundreds of times brighter? Why
SGR1806-20 of 2004 Dec. 27th, shows no evidence of the loss of its
period $P$ or its derivative $\dot{P}$ after the huge
\textit{Magnetar} eruption, while in this model its hypothetical
magnetic energy reservoir (linearly proportional to
$P\cdot\dot{P}$) must be largely exhausted? Why do SGR1806 radio
afterglows show a mysterious two-bump radio curve implying
additional energy injection many days later? In this connection
why are the GRB021004 light curves (from X to radio) calling for
an early and late energy injection? Why has the SGR1806-20
polarization curve been changing angle radically in short ($\sim$
days) timescale? Why is the short GRB050724 able to bump and
re-bright a day after the main burst\cite{Campana}? Why rarest
GRB940217, highest energetic event, could held more than $5000$s.?
\begin{figure}[t]
\begin{center}
\includegraphics[width=3.2in]{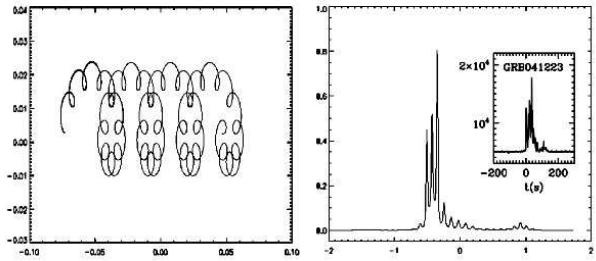}
\caption{The possible simple beam track of a precessing jet to
observer located at origin. On the left, observer stays in (0.00 ;
0.00); the progenitor electron pair jet (leading by IC\cite{FaSa98}
to a gamma jet) has here a Lorentz factor of a thousand and
consequent solid angle at $\sim\mu$ sr. Its consequent blazing light
curve corresponding to such a similar outcome observed in
GRB041223.} \label{fig3}
\end{center}
\end{figure}

Once these major questions are addressed and (in our opinion) mostly
solved by our precessing gamma jet model, a final  question still
remains, calling for a radical assumption on the thin precessing
gamma jet: how can an ultra-relativistic electron beam (in any kind
of Jet models) survive the SN background and dense matter layers and
escape in the outer space while remaining collimated? Such questions
are ignored in most Fireball models that try to fit the very
different GRB afterglow light curves with  shock waves on tuned
shells and polynomial ad-hoc curves around the GRB event. Their
solution forces us more and more toward a unified  precessing Gamma
Jet model feeded by the PeV-TeV lepton showering (about UHE
showering beam see analogous ones\cite{Fa97, Fa00-04}) into
$\gamma$ discussed below. As we will show, the thin gamma precessing
jet is indeed made by a chain of primary processes (PeV muon pair
bundles decaying into electrons and then radiating via synchrotron
radiation), requiring an inner  ultra-relativistic jet inside the
source.
\begin{figure}[t]
\begin{center}
\includegraphics[width=3.2in]{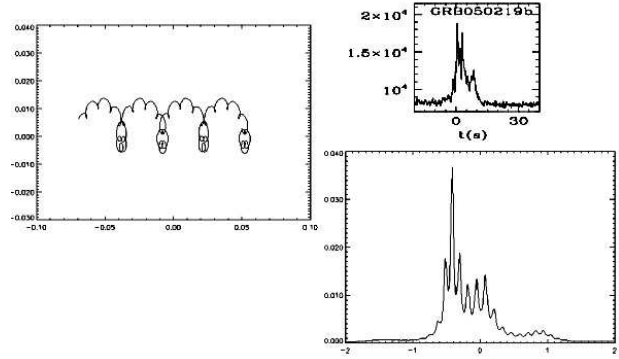}
\caption{Same as in Fig. \ref{fig3}: a precessing jet and its
consequent light curve versus a similar outcome observed in
GRB050219b.} \label{fig4}
\end{center}
\end{figure}
\begin{figure}[h]
\begin{center}
\includegraphics[width=1.8in]{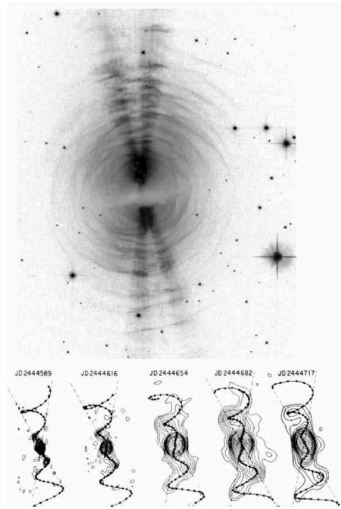}
\caption{The Egg Nebula whose shape might be explained as the
conical section of a twin precessing jet interacting with the
surrounding ejected gas cloud. \emph{Down}: The similar observed
structure of the outflows from the microquasar SS433. A kinematic
model of the time evolution of two oppositely directed precessing
jets is overlaid on the radio contours (\cite{BB}).}\label{fig5}
\end{center}
\end{figure}

\section{Blazing Precessing jets in GRBs and SGRs}
The huge GRBs luminosity (up to $10^{54}$ erg s$^{-1}$) may be due
to a high collimated on-axis blazing jet, powered by a Supernova
output; the gamma jet is made by relativistic synchrotron radiation
and the inner the jet the harder and the denser is its output. The
harder the photon energy, the thinner is the jet opening angle
\begin{figure}[t]
\begin{center}
\includegraphics[width=1.9in,height=2.0in]{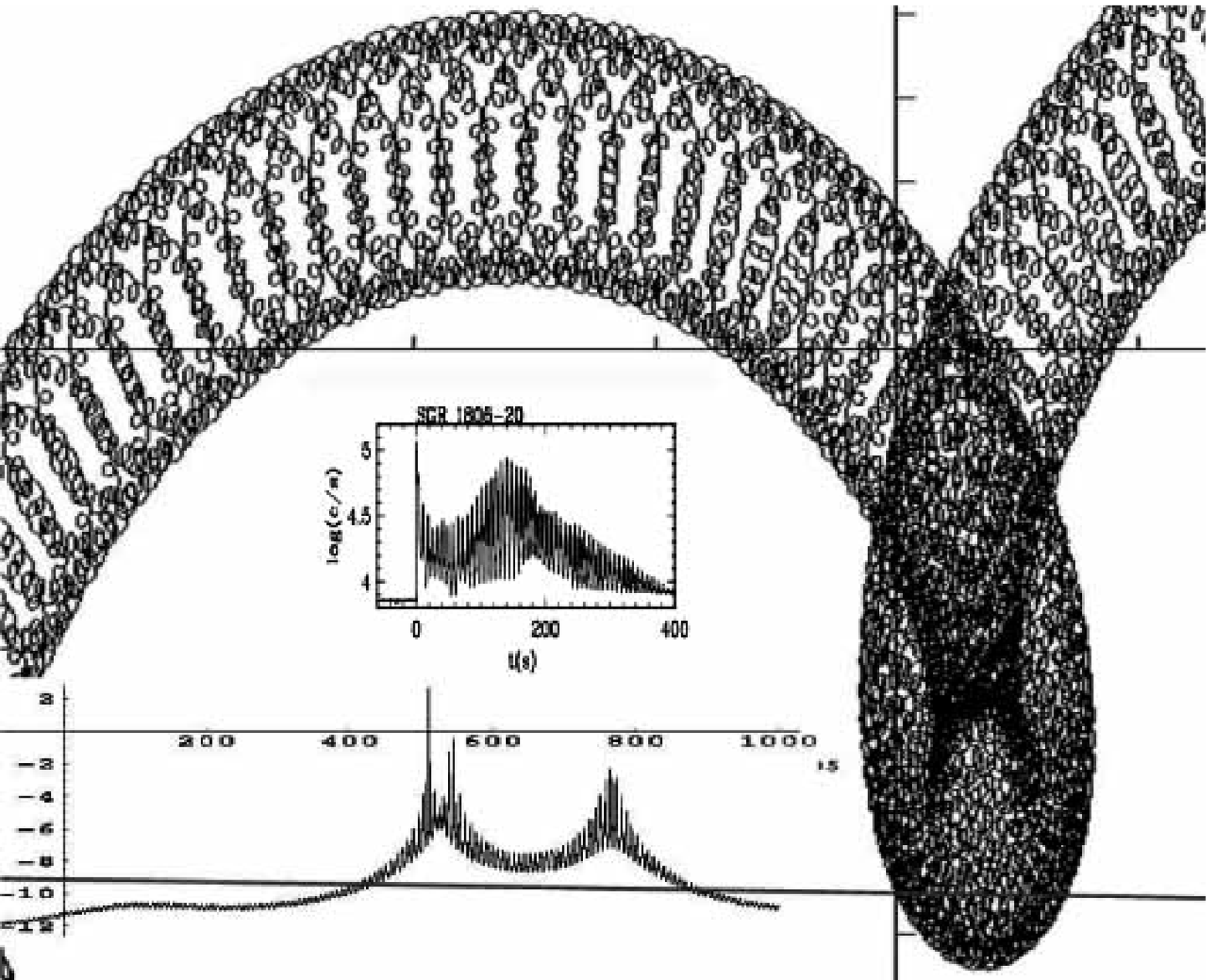}
\includegraphics[width=1.9in,height=2.0in]{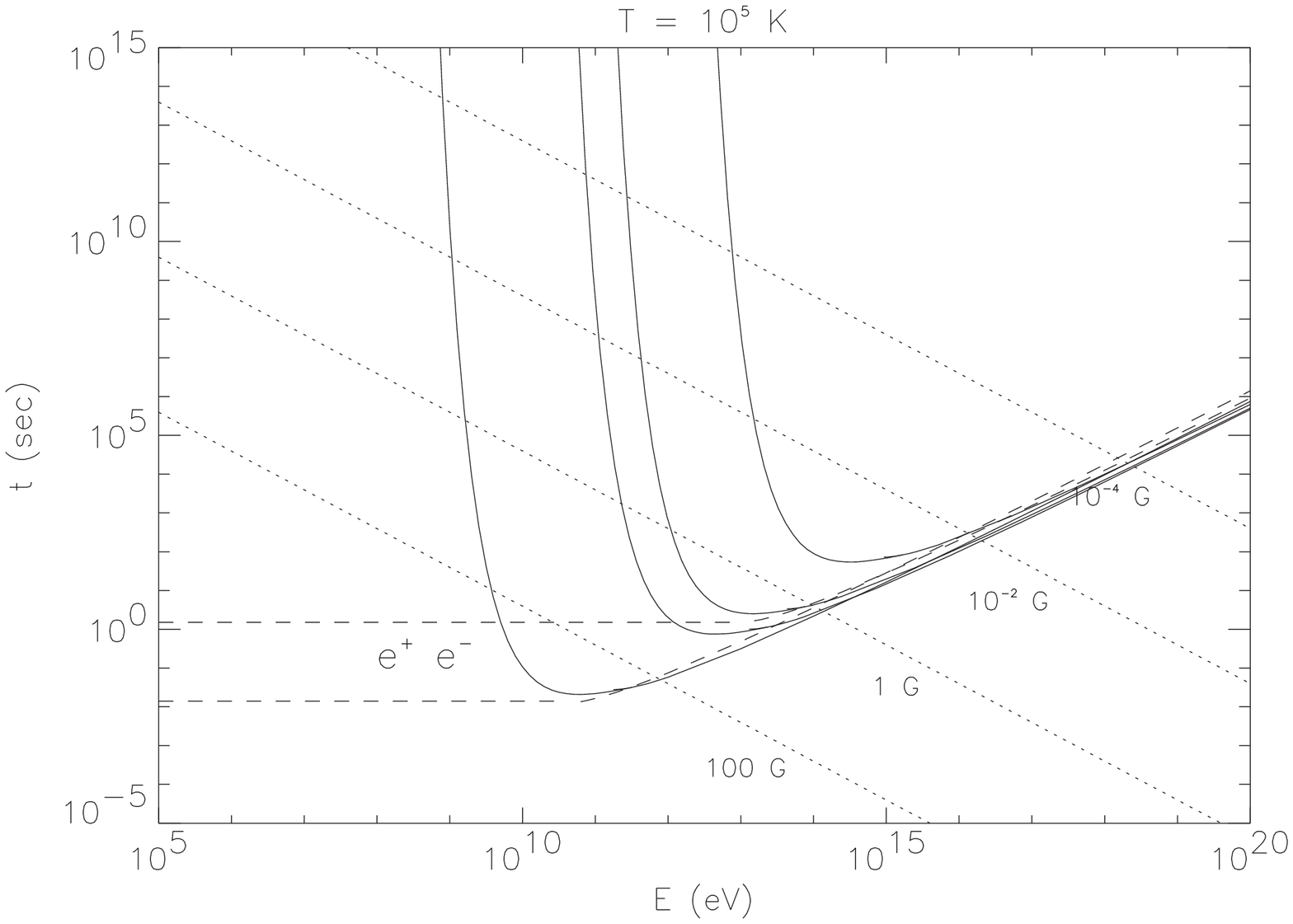}
\caption{\textit{Left}: The more structured multi precessing jet
able to describe the most rare blazing and oscillating SGR 1806-20.
The inner spirals are reflecting the precursor trembling while the
Jet lorentz factor is assumed here at $\gamma_e\simeq10^9$, but in
general the jet showering structure is not cone-like but fan-like
because external stellar magnetic field presence\cite{Fa97,
Fa00-04}, where $\theta\simeq\gamma_e^{-1}$. \emph{Right}:The
electron and muon interaction lengths. The {\em dashed-dotted} and
{\em dotted} lines correspond to the synchrotron energy loss
distance (for muons and electrons respectively) for different values
of the magnetic field: 100 G, 1 G and $10^{-2}$ G. The {\em straight
solid} line labelled $t_{\mu}$ indicates the muon lifetime; the {\em
dashed} lines indicate the IC interaction lengths for muons and
electrons. Finally the two {\em solid} curves labeled $\mu^+ \mu^-$
and $e^+ e^-$ correspond to the attenuation length of high energy
photons producing lepton pairs (either $\mu^{\pm}$, or $e^{\pm}$)
through the interaction with the SN radiation field. We have assumed
that the thermal photons emitted by the star in a pre-SN phase have
a black body distribution with a temperature $T\simeq10^5$ K.
Assuming a radius $R\sim10\,R_{\odot}$, we are considering a
luminosity of $L_{SN}\simeq2.5\cdot10^{41}$ erg $s^{-1}$. Around
$10^{15}$-$10^{16}$ eV, muons decay before losing energy via IC
scattering with the stellar background or via synchrotron radiation.
Right: The Supernova opacity (interaction length) for PeV electrons
at different times. PeV muon jets may overcome it and decay later in
$\gamma$ showering electrons (see for details
\cite{DaF05}).}\label{fig6}
\end{center}
\end{figure}
$\Delta\theta\simeq\gamma^{-1}$, $\Delta\Omega\simeq\gamma^{-2}$
where $\gamma\simeq10^4$. The thin solid angle explains the rare
SN-GRB connection and for instance the apparent GRB990123
extraordinary power (billions of times the typical SN luminosity).
This also explains the rarer, because nearer, GRB-SN events such as
GRB980425 or GRB060218, whose jets were off-axis
$\simeq300\cdot\gamma^{-1}$, i.e. a few degrees, increasing its
detection probability by roughly a hundred thousand times, but whose
GRB luminosity was almost by the same factor extremely low. This
beaming selection in larger volumes explains the puzzling evidence
(the Amati correlation) of harder and apparently powerful GRBs at
larger and larger distances in opposition to Hubble law calling for
redder and redder signals. The statistical selection favors (in
wider volumes and for a wider sample of SN-GRB-jet) the harder and
more on-axis events. A huge (million time) unobserved population of
far off-axis SN-GRBs at cosmic distances are below the detection
thresholds. This Amati correlation remains unexplained for any
isotropic or fountain cone Fireball model and it is in contrast with
the cosmic trend required by the Hubble-Friedmann law: the further
the distances, the larger the redshifts, the smoother the time lag
profile and the softer the expected GRB event.
\begin{figure}[t]
\begin{center}
\includegraphics[width=1.8in]{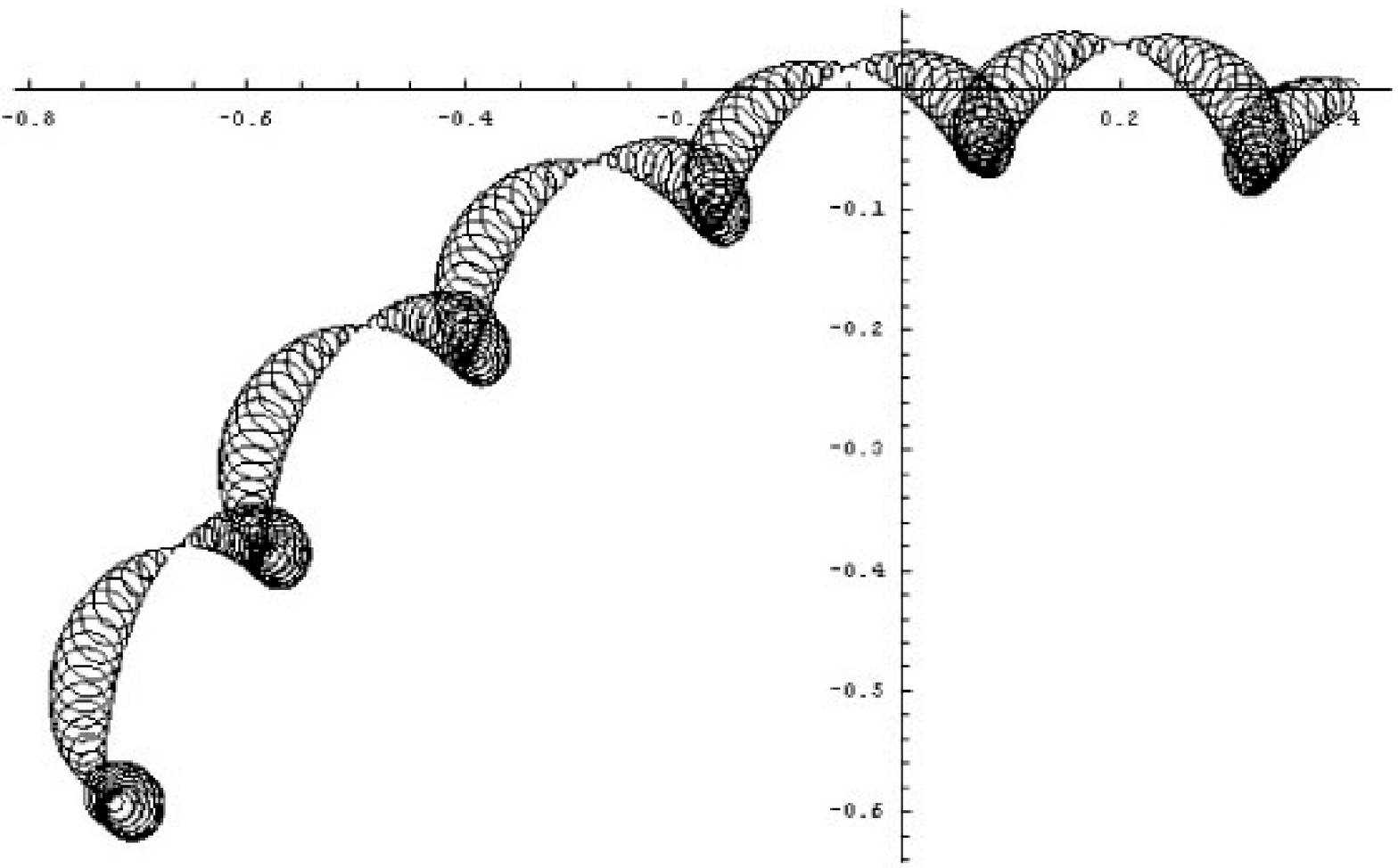}
\includegraphics[width=1.8in]{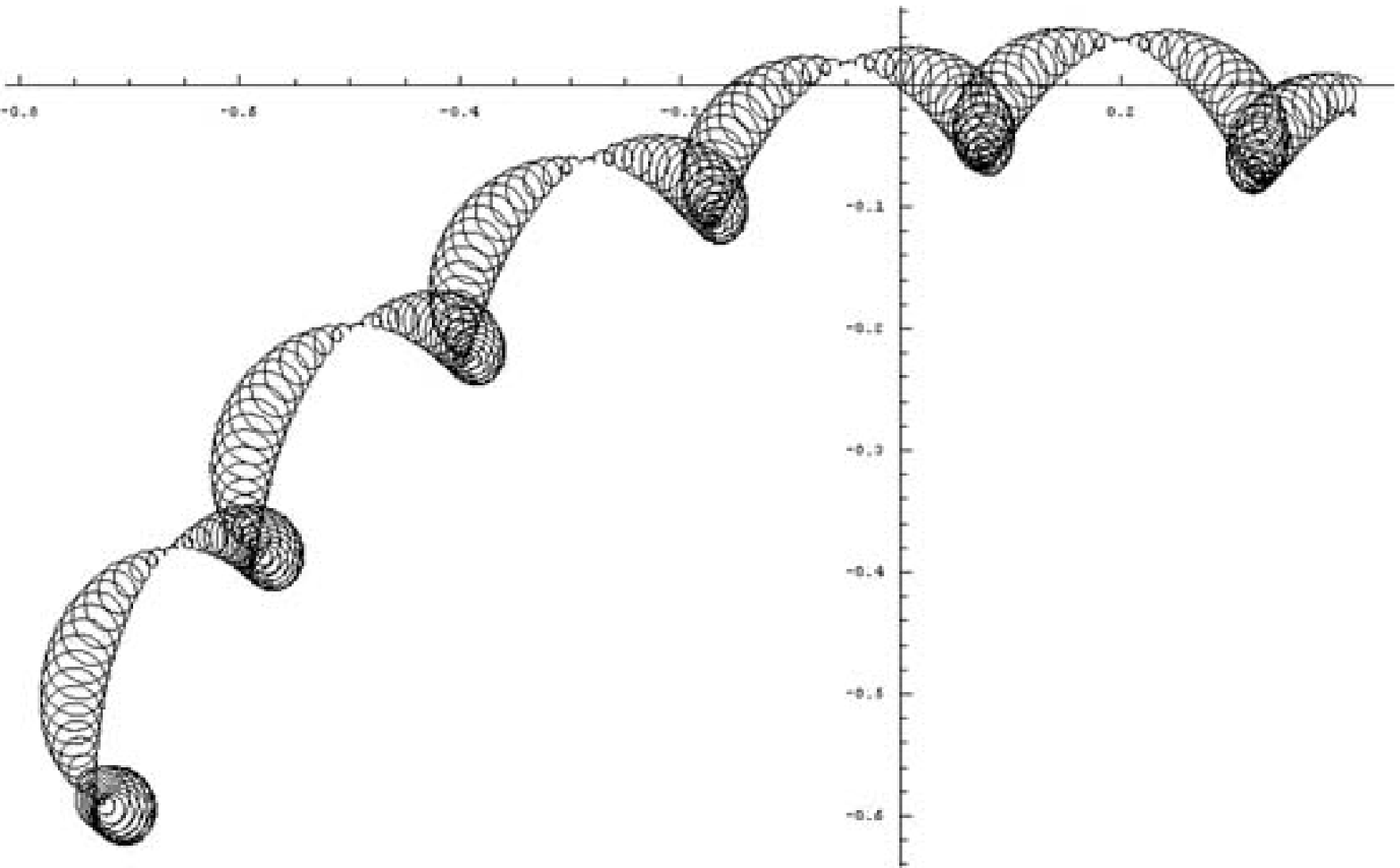}
\includegraphics[width=1.8in]{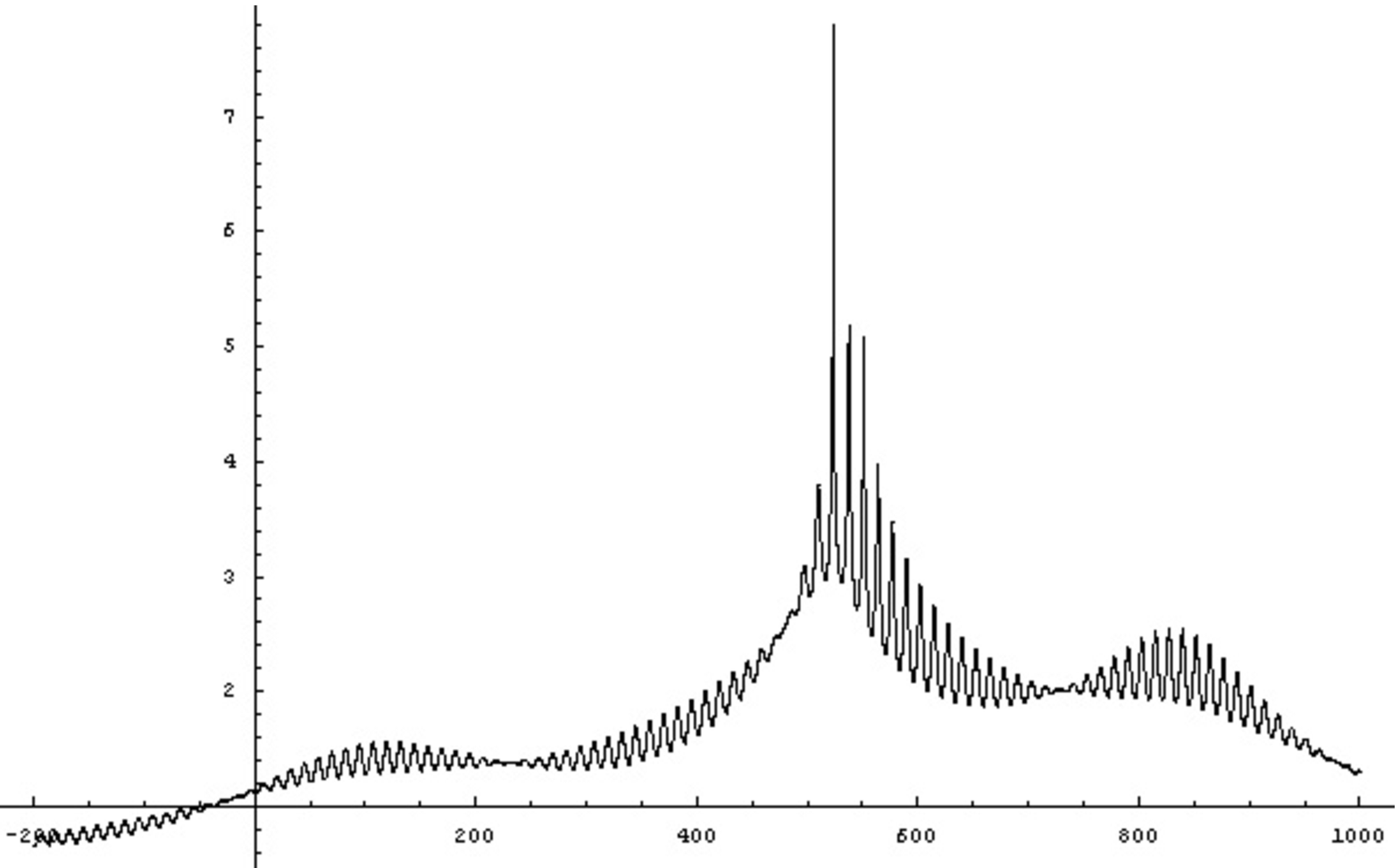}
\includegraphics[width=1.8in]{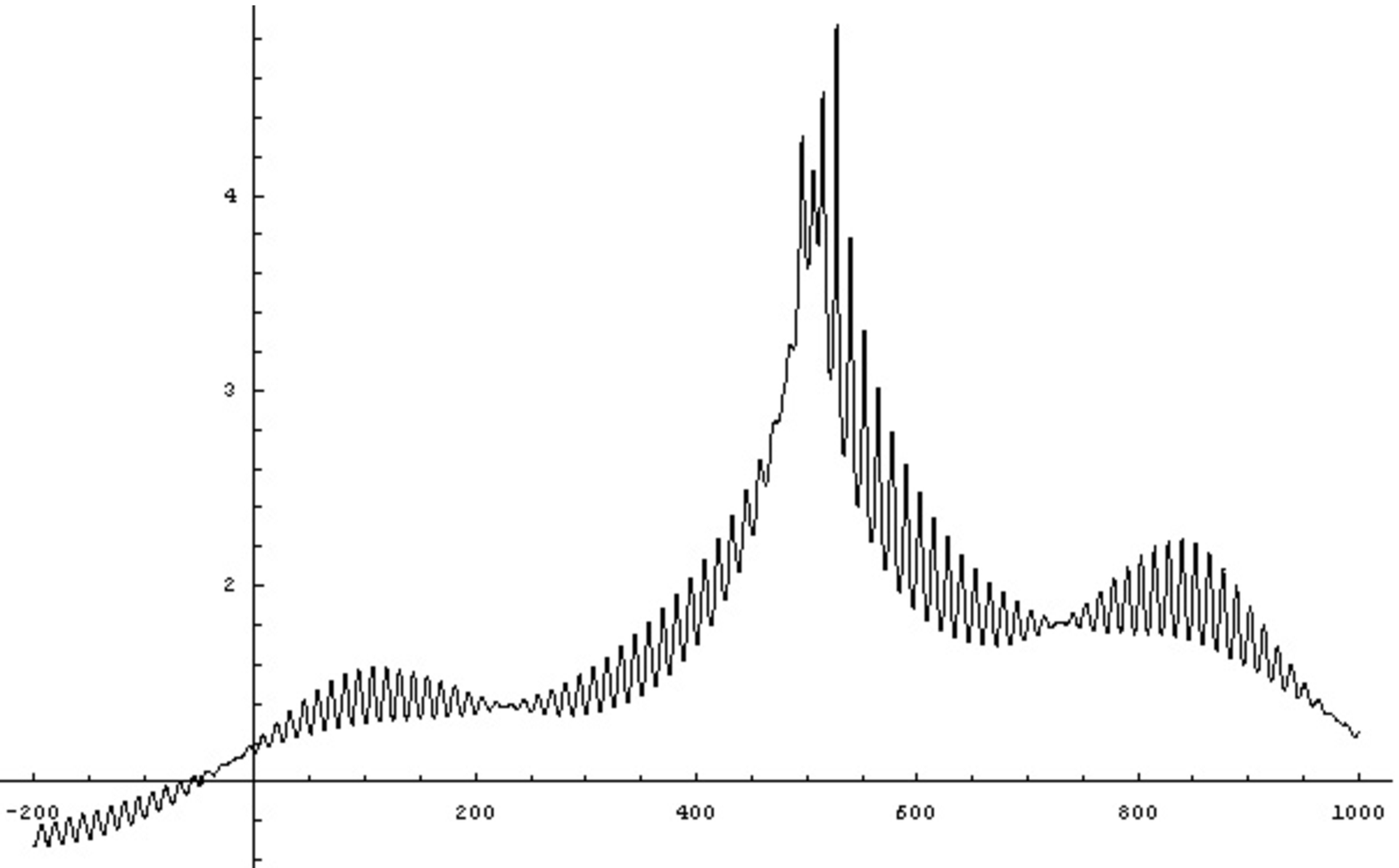}
\caption{\textit{Left and right}: A more beamed and a less beamed
(non linear) precessing blazing; note  on the right side that a
much off-axis beaming induce a different SGR or GRB smoother and
softer profile and a much limited  amplification.} \label{fig7}
\end{center}
\end{figure}
In our model to make GRB-SN in nearly energy equipartition the jet
must be very collimated $\frac{\Omega}{\Delta\Omega}\simeq
10^{8}$-$10^{10}$ (\cite{FaSa95b, Fa99, DaF05}) explaining why
apparent (but beamed) GRB luminosity $\dot{E}_{GR-jet}\simeq
10^{53}$-$10^{54}$ erg $s^{-1}$ coexist on the same place and
similar epochs with lower (isotropic) SN powers $\dot{E}_{SN}\simeq
10^{44}-10^{45} erg s^{-1}$. In order to fit the statistics between
GRB-SN rates, the jet must have a decaying activity ($\dot{L}\simeq
(\frac{t}{t_o})^{-\alpha}$, $\alpha \simeq 1$): it must survive not
just for the observed GRB duration but for a much longer timescale,
possibly thousands of time longer $t_o\simeq10^4\,s$. The late
stages of the GRBs (within the same decaying power law) would appear
as a SGRs. Indeed the puzzle (for one shot popular Magnetar-Fireball
model\cite{DuTh92}) arises for the surprising giant flare from SGR
1806-20 that occurred on 2004 December 27th: if it has been radiated
isotropically (as assumed by the Magnetar model\cite{DuTh92}), most
of - if not all - the magnetic energy stored in the neutron star NS,
should have been consumed at once. This should have been reflected
into sudden angular velocity loss (and-or its derivative) which was
\textit{never observed}. On the contrary a thin collimated
precessing jet $\dot{E}_{SGR-jet}\simeq 10^{36}$-$10^{38}$ erg
$s^{-1}$, blazing on-axis, may be the source of such an apparently
(the inverse of the solid beam angle
$\frac{\Omega}{\Delta\Omega}\simeq10^{8}$-$10^{9}$) huge bursts
$\dot{E}_{SGR-Flare}\simeq10^{38}\cdot\frac{\Omega}{\Delta\Omega}\simeq10^{47}$
erg $s^{-1}$ with a moderate  steady jet output power (X-Pulsar,
SS433). This explains the absence of any variation in the SGR1806-20
period and its time derivative, contrary to any obvious correlation
with the dipole energy loss law. The nearby spiralling of the jet to
us explains the later bumping oscillatory signal of the event (see
Fig. \ref{fig7}).
\begin{figure}[t]
\begin{center}
\includegraphics[width=2.2in]{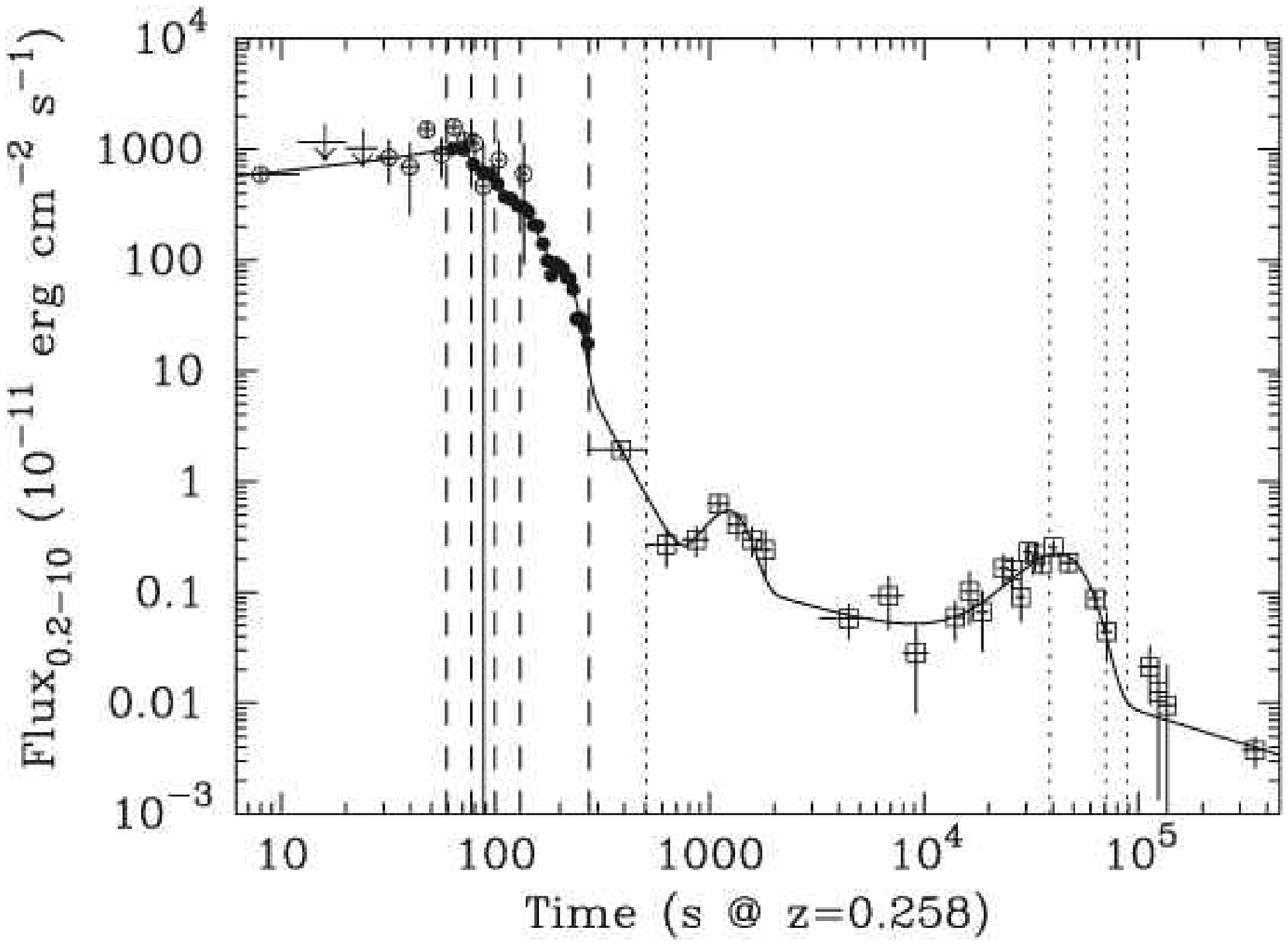}
\includegraphics[width=1.9in]{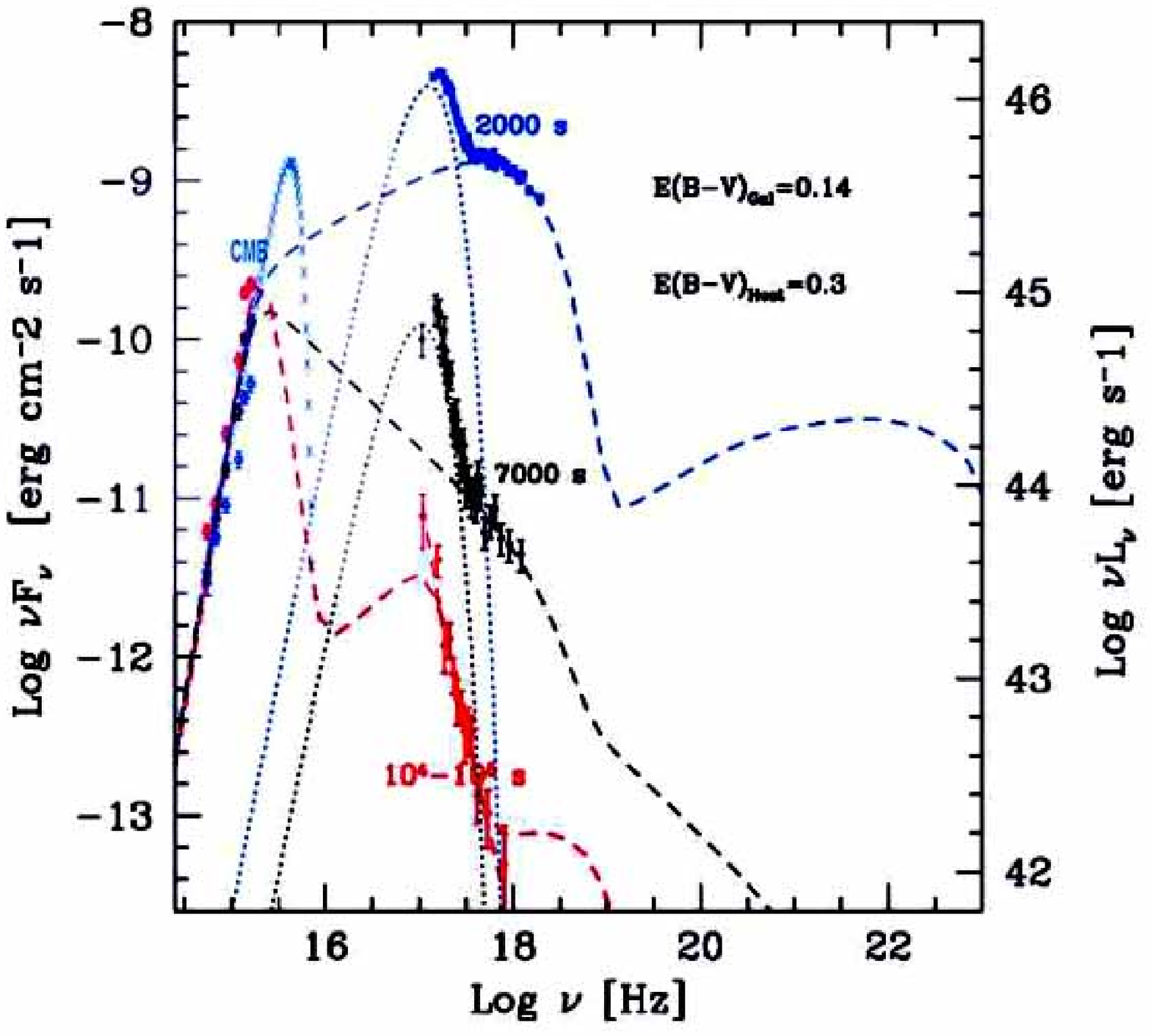}
\caption{Left: The short GRB050724 and its long life X-ray afterglow
whose curve (\cite{Campana}) and whose multi-re-brightening is
testing the persistent jet activity and  geometrical blazing views.
Right: the very recent optical afterglows of GRB 060218  whose
smooth longest X-ray flare coexist with a thermal Black Body
Radiation (see previous figure)  component (a steady SN bump) while
the external Jet cone tail are fading while pointing off-axis
elsewhere, adapted from \cite{Ghisellini, Moretti,
Fargion-GNC}).}\label{fig8}
\end{center}
\end{figure}

In our model, the temporal evolution of the angle between the
spinnng (PSRs), precessing (binary, nutating) jet direction and the
rotational axis of the NS, can be expressed as
\[
\theta_1(t)=\sqrt{\theta_x^2+\theta_y^2}
\]
where
$$\theta_y(t)=
\theta_a\cdot\sin\omega_0t+\cos(\omega_bt+\phi_{b})+\theta_{psr}\cdot\cos(\omega_{psr}t+\phi_{psr})\cdot|(\sin(\omega_Nt+\phi_N))|+
$$
$$
+\theta_s\cdot\cos(\omega_st+\phi_{s})+\theta_N\cdot\cos(\omega_Nt+\phi_N))+\theta_y(0)
$$
and a similar law express the $\theta_x(t)$ evolution. The angular
velocities and phase labels are self-explained\cite{DaF05, DaF06}.
Lorentz factor $\gamma$ of the jet's relativistic particles, for the
most powerful SGR1806-20 event, and other parameters adopted for the
jet model represented in Fig. \ref{fig2} are shown in the following
Table \ref{Tab1} (\cite{DaF05, DaF06}).
\begin{table}
\begin{center}
\begin{tabular}{lll}
\hline \hline
  $\gamma = 10^9$  & $\theta_a=0.2$ & $\omega_a =1.6 \cdot 10^{-8}$ rad/s\\
  $\theta_b=1$ &  $\theta_{psr}$=1.5 $\cdot 10^7$/$\gamma$ & $\theta_N$=$5 \cdot 10^7$/$\gamma$ \\
$\omega_b$=4.9 $\cdot 10^{-4}$ rad/s &  $\omega_{psr}$=0.83 rad/s
& $\omega_N $=1.38 $\cdot 10^{-2}$ rad/s \\
$\phi_{b}=2\pi - 0.44$ &$\phi_{psr}$=$\pi + \pi/4$ & $\phi_N$=3.5
$\pi/2 + \pi/3$ \\
$\phi_s \sim \phi_{psr}$ & $\theta_s$=1.5 $\cdot 10^6$/$\gamma$ & $\omega_s = 25$ rad/s \\
 \hline \hline
\end{tabular}
 \label{Tab1}
\end{center}
\end{table}

The simplest  way to produce the $\gamma$ emission  would be by IC
of GeVs electron pairs onto thermal infra-red photons. Also
electromagnetic showering of PeV electron pairs by synchrotron
emission in galactic fields, ($e^{\pm}$ from muon decay) may be the
progenitor of the $\gamma$ blazing jet. However, the main difficulty
for a jet of GeV electrons is that their propagation through the SN
radiation field is highly suppressed. UHE muons
($E_{\mu}\gtrsim$~PeV) instead are characterized by a longer
interaction length either with the circum-stellar matter and the
radiation field, thus they have the advantage to avoid the opacity
of the star and escape the dense GRB-SN isotropic radiation field
\cite{DaF05, DaF06}. We propose that also the emission of SGRs is
due to a primary hadronic jet producing ultra relativistic $e^{\pm}$
(1 - 10 PeV) from hundreds PeV pions, $\pi\rightarrow\mu\rightarrow
e$, (as well as EeV neutron decay in flight): primary protons can be
accelerated by the large magnetic field of the NS up to EeV energy.
The protons could in principle emit directly soft gamma rays via
synchrotron radiation with the galactic magnetic field
($E_{\gamma}^p\simeq10(E_p/EeV)^2(B/2.5\cdot10^{-6}\,G)$ keV), but
the efficiency is poor because of the too small proton
cross-section, too long timescale of proton synchrotron
interactions. By interacting with the local galactic magnetic field
relativistic pair electrons lose energy via synchrotron radiation:
$E_{\gamma}^{sync}\simeq4.2\cdot10^6(\frac{E_e}{5\cdot10^{15}\,eV})^2(\frac{B}{2.5\cdot10^{-6}\,G})\,eV$
with a characteristic timescale
$t^{sync}\simeq1.3\cdot10^{10}(\frac{E_{e}}{5\cdot10^{15}\,eV})^{-1}(\frac{B}{2.5\cdot10^{-6}\,G})^{-2}\,s$.
This mechanism would produce a few hundreds keV radiation as it is
observed in the intense $\gamma$-ray flare from SGR 1806-20.

The Larmor radius is about two orders of magnitude smaller than the
synchrotron interaction length and this may imply that the aperture
of the showering jet is spread in a fan structure \cite{Fa97,
Fa00-04} by the magnetic field,
$\frac{R_L}{c}\simeq4.1\cdot10^{8}(\frac{E_{e}}{5\cdot10^{15}\,eV})(\frac{B}{2.5\cdot10^{-6}\,G})^{-1}\,s$.
Therefore the solid angle is here the inverse of the Lorentz factor
($\sim$ nsr). In particular a thin
($\Delta\Omega\simeq10^{-9}$-$10^{-10}$ sr) precessing jet from a
pulsar may naturally explain the negligible variation of the spin
frequency $\nu=1/P$ after the giant flare ($\Delta\nu<10^{-5}$ Hz).
Indeed it seems quite unlucky that a huge
($E_{Flare}\simeq5\cdot10^{46}$ erg) explosive event, as the needed
mini-fireball by a magnetar model\cite{DuTh92}, is not leaving any
trace in the rotational energy of the SGR 1806-20, $
E_{rot}=\frac{1}{2}I_{NS}\omega^2\simeq3.6\cdot10^{44}(\frac{P}{7.5\,s})^{-2}(\frac{I_{NS}}{10^{45}g\,cm^2})$
erg. The consequent fraction of energy lost after the flare is
severely bounded by observations:
$\frac{\Delta(E_{Rot})}{E_{Flare}}\leq10^{-6}$. More absurd in
Magnetar-explosive model is the evidence of a brief precursor event
(one-second SN output) taking place with no disturbance on
SGR1806-20 \textit{two minutes before} the hugest flare of 2004 Dec.
27th. The thin precessing Jet while being extremely collimated
(solid angle $\frac{\Omega}{\Delta\Omega}\simeq10^{8}$-$10^{10}$
(\cite{FaSa95b, Fa99, DaF05, DaF06}) may blaze at different angles
within a wide energy range (inverse of
$\frac{\Omega}{\Delta\Omega}\simeq10^{8}$-$10^{10}$). The output
power may exceed $\simeq10^{8}$, explaining the extreme low observed
output in GRB980425 -an off-axis event-, the long late off-axis
gamma tail by  GRB060218\cite{Fargion-GNC}),  respect to the on-axis
and more distant GRB990123 (as well as GRB050904). \textbf{In
conclusion \textit{GRBs are not the most powerful explosions, but
just the most  collimated ones.}}

\end{document}